# Experimental Analysis of AODV & DSR over TCP & CBR Connections with Varying Speed and Node Density in VANET

Bijan Paul
Dept. of Computer Science & Engineering,
Shahjalal University of Science & Technology,
Sylhet, Bangladesh

Md. Ibrahim
Dept. of Computer Science & Engineering,
Shahjalal University of Science & Technology,
Sylhet, Bangladesh

Md. Abu Naser Bikas
Lecturer, Dept. of Computer Science & Engineering,
Shahjalal University of Science & Technology,
Sylhet, Bangladesh

## ABSTRACT
Vehicular adhoc network or **VANET** is special types of adhoc network consists of moving cars referred to as nodes; provide a way to exchange any information between cars without depending on fixed infrastructure. For efficient communication between nodes various routing protocols and mobility models have been proposed based on different scenarios. Due to rapid topology changing and frequent disconnection makes it difficult to select suitable mobility model and routing protocols. Hence performance evaluation and comparison between routing protocols is required to understand any routing protocol as well as to develop a new routing protocol. In this research paper, the performance of two on-demand routing protocols AODV & DSR has been analyzed by means of packet delivery ratio, loss packet ratio & average end-to-end delay with varying speed limit and node density under TCP & CBR connection.

## 1. INTRODUCTION
VANET (vehicular adhoc network) is a autonomous & self-organizing wireless communication network .In this network the cars are called nodes which involve themselves as servers and/or clients for exchanging & sharing information. This is a new technology thus government has taken huge attention on it. There are many research projects around the world which are related with VANET such as COMCAR [1], DRIVE [2], FleetNet [3] and NoW (Network on Wheels) [4], CarTALK 2000 [5], CarNet [6].
There are several VANET applications such as Vehicle collision warning, Security distance warning, Driver assistance, Cooperative driving, Cooperative cruise control, Dissemination of road information, Internet access, Map location, Automatic parking, Driverless vehicles.
In this paper, we have evaluated performance of AODV and DSR based on TCP and CBR connection with varying speed time and also various network parameters and measured performance metrics such as packet delivery
ratio, loss packet ratio and average end-to-end delay of this two routing protocol and compared their performance. The remainder of the paper is organized as follows: Section 2 describes two unicast routing protocols AODV and DSR of VANET. Section 3 describes connection types like TCP and CBR. Section 4 presents performance metrics and the network parameters. Section 5 presents our implementation. We conclude in Section 6 and section 7 for reference.

## 2. ROUTING PROTOCOLS
An ad hoc routing protocol [7] is a convention, or standard, that controls how nodes decide which way to route packets in between computing devices in a mobile adhoc network.
The routing protocol of VANET can be classified into two categories such as Topology based routing protocols & Position based routing protocols. Existing unicast routing protocols of VANET is not capable to meet every traffic scenarios. They have some pros and cons. We have already described it in our previous work [8].For our simulation purpose we have selected two on demand routing protocols AODV & DSR.

### 2.1 AODV
Ad Hoc on Demand Distance Vector routing protocol [9] is a reactive routing protocol which establish a route when a node requires sending data packets. It has the ability of unicast & multicast routing. It uses a destination sequence number (DestSeqNum) which makes it different from other on demand routing protocols. It maintains routing tables, one entry per destination and an entry is discarded if it is not used recently. It establishes route by using RREQ and RREP cycle. If any link failure occurs, it sends report and another RREQ is made.

### 2.2 DSR
The Dynamic Source Routing (DSR) [10] protocol utilizes source routing & maintains active routes. It has two phases route discovery & route maintenance. It does not use periodic routing message. It will generate an error message if there is any link failure. All the intermediate nodes ID are stored in the packet header of DSR. If there has multiple paths to go to the destination DSR stores multiple path of its routing information.
AODV and DSR have some significant differences. In AODV when a node sends a packet to the destination then data packets only contains destination address. On the other hand in DSR when a node sends a packet to the destination the full routing information is carried by data packets which causes more routing overhead than AODV.

## 3. Connection Types

There are several types of connection pattern in VANET. For our simulation purpose we have used CBR and TCP connection pattern.

### 3.1 Constant Bit Rate (CBR)

Constant bit rate means consistent bits rate in traffic are supplied to the network. In CBR, data packets are sent with fixed size and fixed interval between each data packets. Establishment phase of connection between nodes is not required here, even the receiving node don't send any acknowledgement messages. Connection is one way direction like source to destination.

### 3.2 Transmission Control Protocol (TCP)

TCP is a connection oriented and reliable transport protocol. To ensure reliable data transfer TCP uses acknowledgement, time outs and retransmission. Acknowledge means successful transmission of packets from source to destination. If an acknowledgement is not received during a certain period of time which is called time out then TCP transmit the data again.

## 4. Performance Metrics & Network Parameters

For network simulation, there are several performance metrics which is used to evaluate the performance. In simulation purpose we have used three performance metrics.

### 4.1 Packet Delivery Ratio

Packet delivery ratio is the ratio of number of packets received at the destination to the number of packets sent from the source. The performance is better when packet delivery ratio is high.

### 4.2 Average end-to-end delay

This is the average time delay for data packets from the source node to the destination node. To find out the end-to-end delay the difference of packet sent and received time was stored and then dividing the total time difference over the total number of packet received gave the average end-to-end delay for the received packets. The performance is better when packet end-to-end delay is low.

### 4.3 Loss Packet Ratio (LPR)

Loss Packet Ratio is the ratio of the number of packets that never reached the destination to the number of packets originated by the source.

## 5. OUR IMPLEMENTATION

For simulation purpose we used random waypoint mobility model. Network Simulator NS-2.34[11, 12] has been used. To measure the performance of AODV and DSR we used same scenario for both protocols. Because of both protocols unique behavior the resultant output differ.

### 5.1 Simulation Parameters

In our simulation, we used environment size 840 m x 840 m, node density 30 to 150 nodes with constant pause time 20s and variable speed 5 to 25 m / s. We did the Simulation for 200s with maximum 8 connections. The network parameters we have used for our simulation purpose shown in the table 1.

**Table 1 .Network Parameters**

| Parameter | Value |
|---|---|
| Protocols | AODV, DSR |
| Simulation Time | 200 s |
| Number of Nodes | 30, 60, 90, 120, 150 |
| Simulation Area | 840 m x 840 m |
| Pause Time | 20 s |
| Traffic Type | CBR , TCP |
| Maximum Speed | 5 , 10 , 15 , 20 , 25 m / s |
| Mobility Model | Random Waypoint |
| Network Simulator | NS 2.34 |

### 5.2 Simulation Results

The performance of AODV & DSR has been analyzed with varying speed time 5m/s to 25m/s for number of nodes 30, 60, 90, 120, 150 under TCP & CBR connection. We measure the packet delivery ratio, loss packet ratio & average end-to-end delay of AODV and DSR and the simulated output has shown by using graphs.

### 5.3 Graphs

On left side in module 5.3 we draw the graph of TCP connection simulation result. Similarly, on right side we draw the graph of CBR connection simulation result.

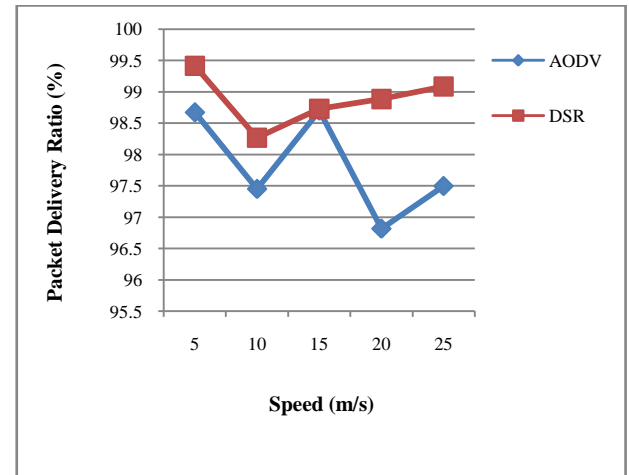

**Fig 1: PDR of 30 nodes using TCP**

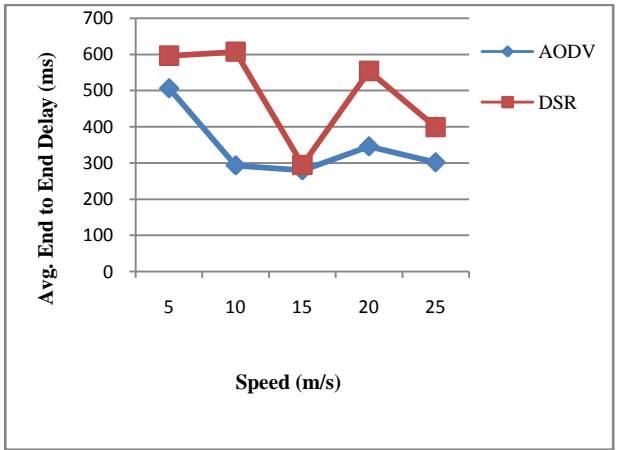
Fig 2: Avg.E-2-E delay of 30 nodes using TCP

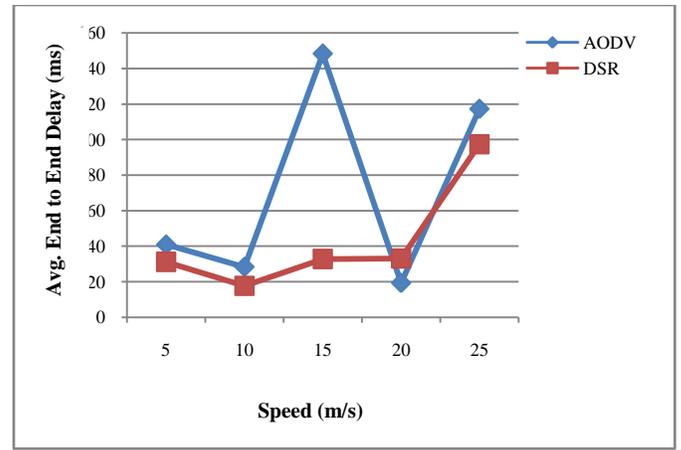
Fig 5: Avg.E-2-E delay of 30 nodes using CBR

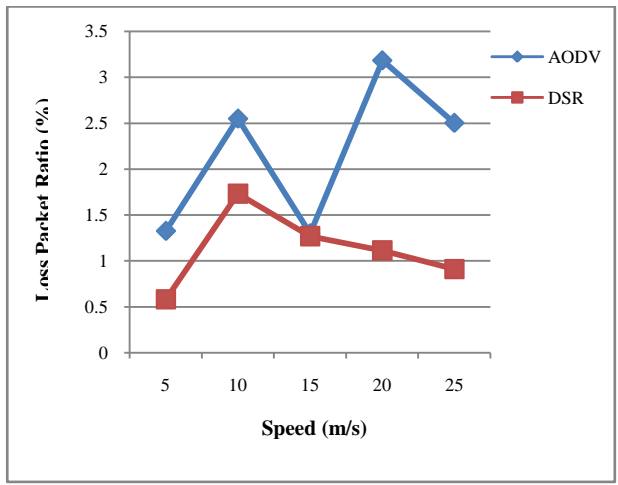
Fig 3: LPR of 30 nodes using TCP

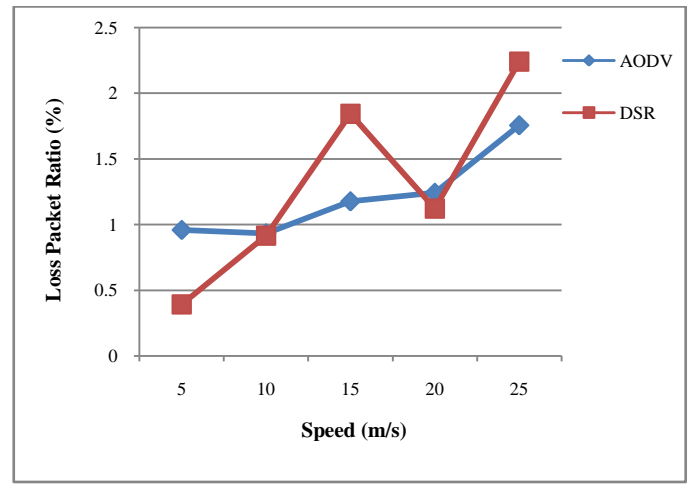
Fig 6: LPR of 30 nodes using CBR

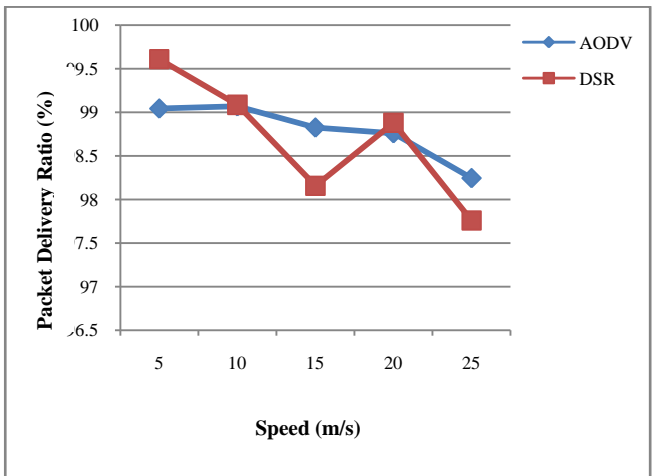
Fig 4: PDR of 30 nodes using CBR

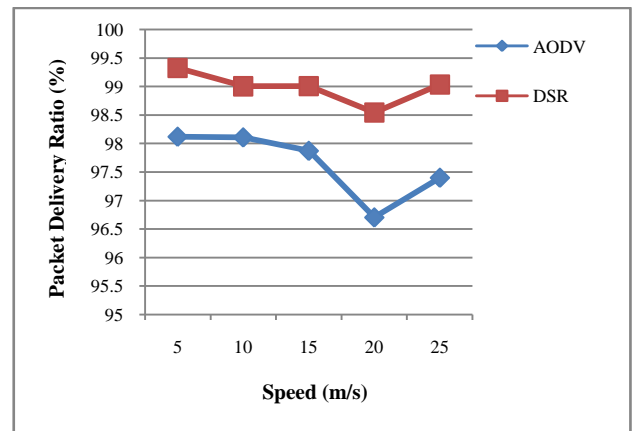
Fig 7: PDR of 60 nodes using TCP

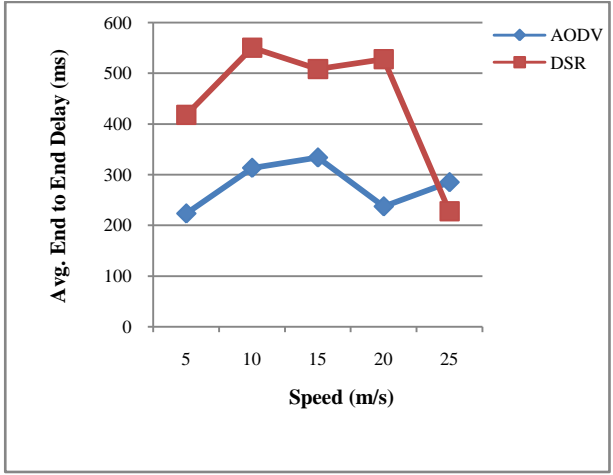

Fig 8: Avg.E-2-E delay of 60 nodes using TCP

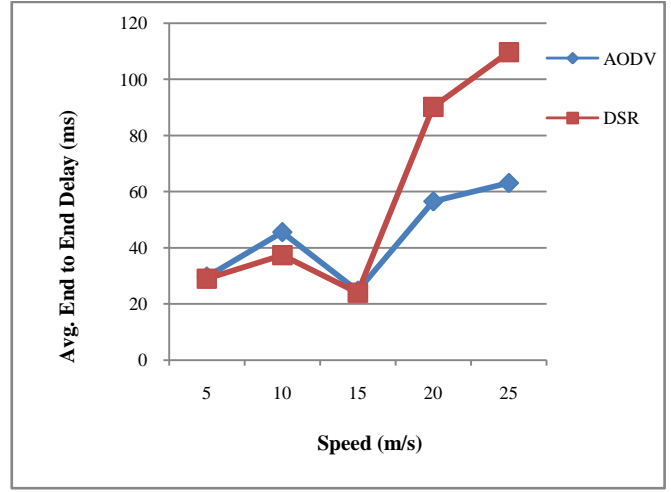

Fig 11: Avg.E-2-E delay of 60 nodes using CBR

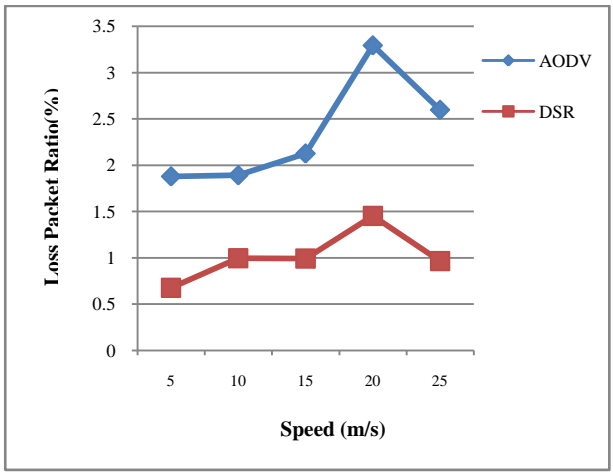

Fig 9: LPR of 60 nodes using TCP

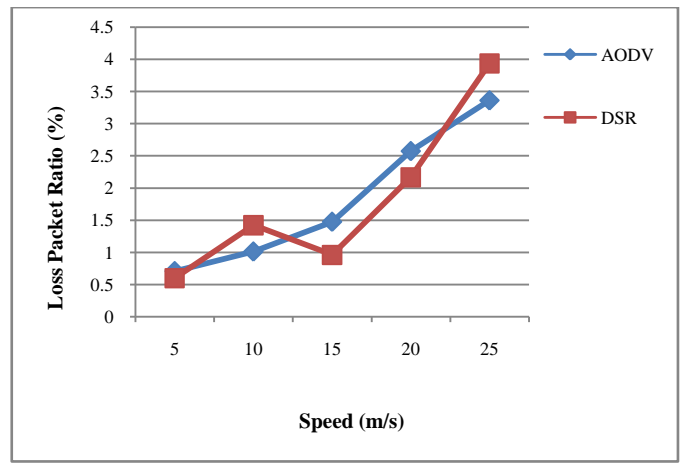

Fig 12: LPR of 60 nodes using CBR

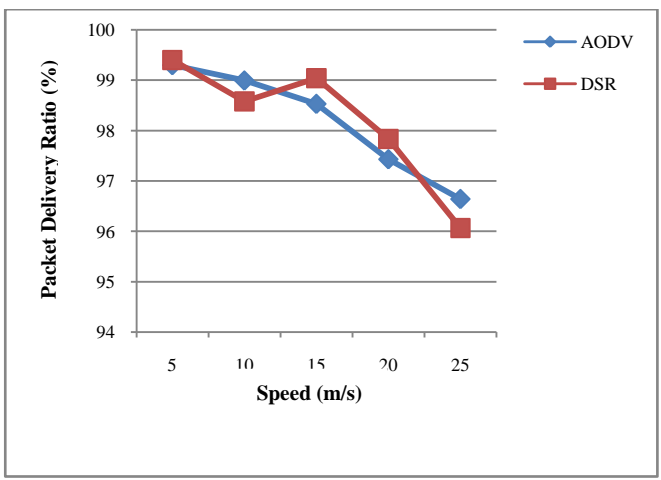

Fig 10: PDR of 60 nodes using CBR

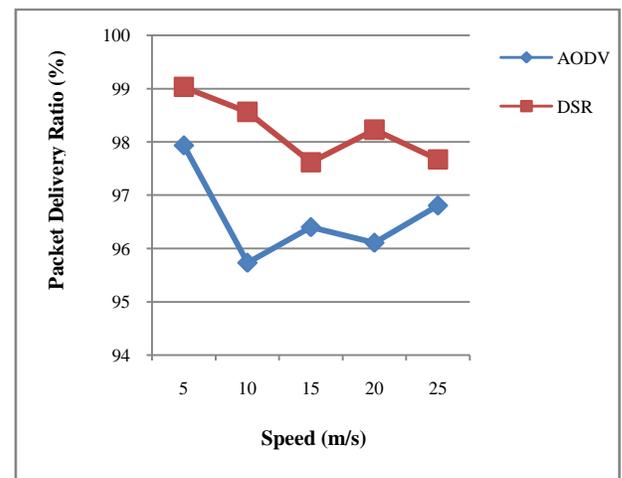

Fig 13: PDR of 90 nodes using TCP

33

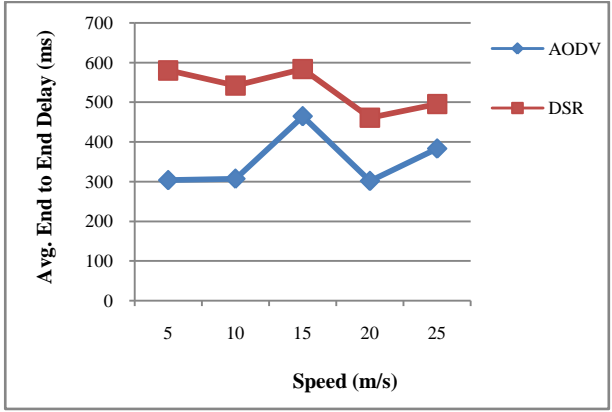

**Fig 14: Avg.E-2-E delay of 90 nodes using TCP**

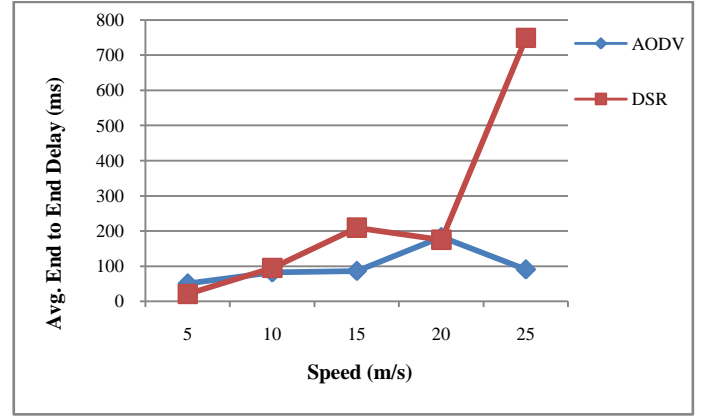

**Fig 17: Avg.E-2-E delay of 90 nodes using CBR**

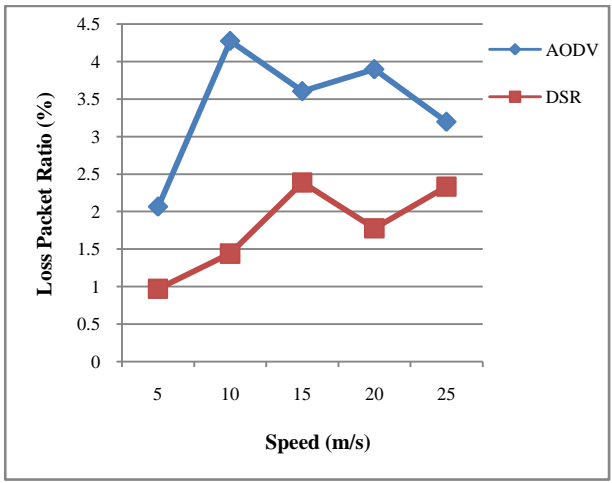

**Fig 15: LPR of 90 nodes using TCP**

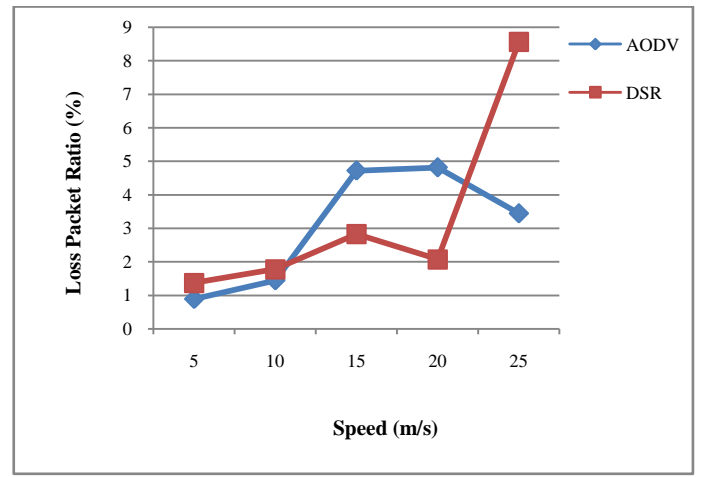

**Fig 18: LPR of 90 nodes using CBR**

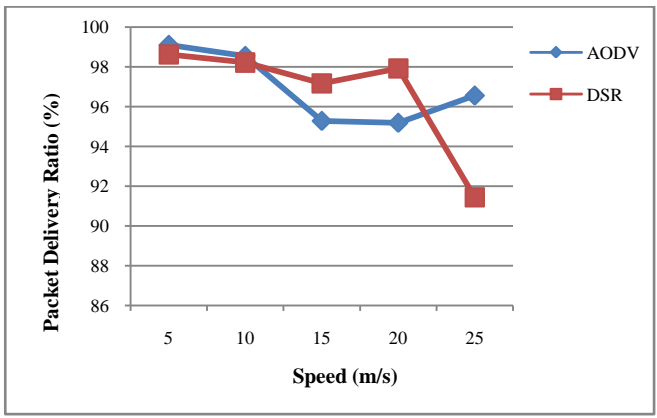

**Fig 16: PDR of 90 nodes using CBR**

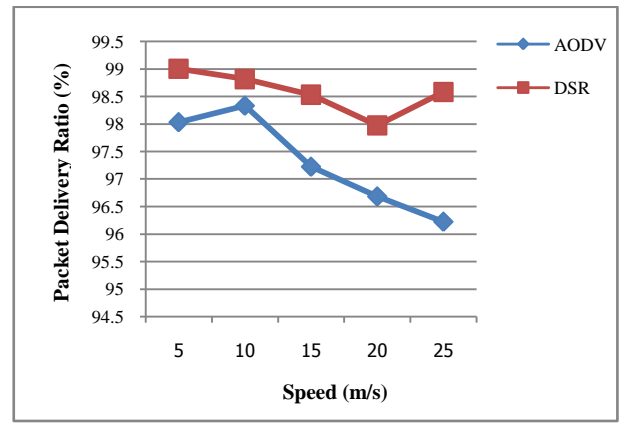

**Fig 19: PDR of 120 nodes using TCP**

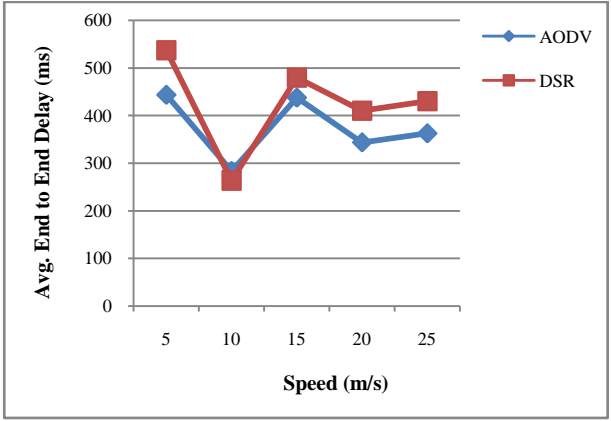

**Fig 20: Avg.E-2-E delay of 120 nodes using TCP**

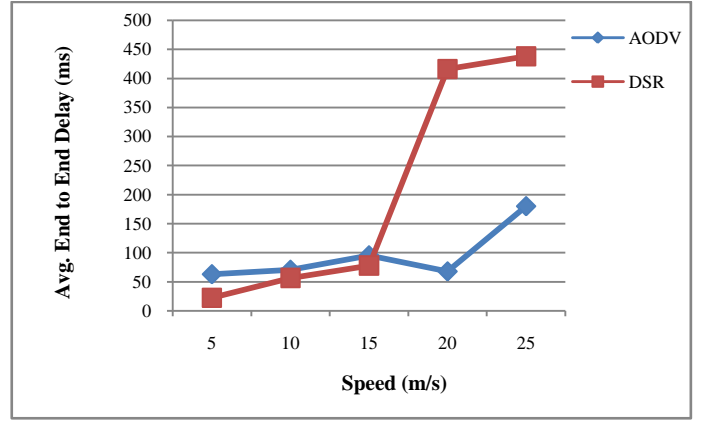

**Fig 23: Avg.E-2-E delay of 120 nodes using CBR**

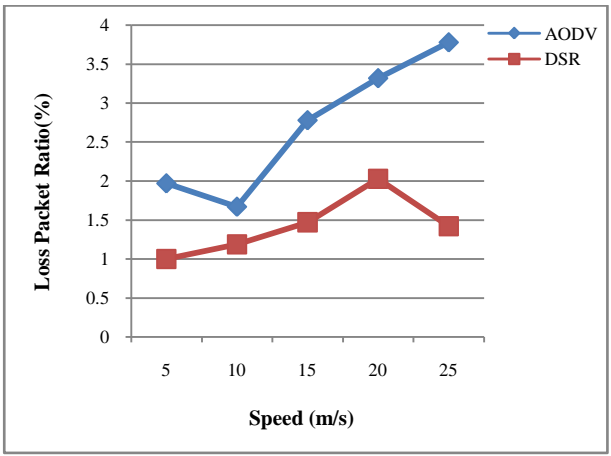

**Fig 21: LPR of 120 nodes using TCP**

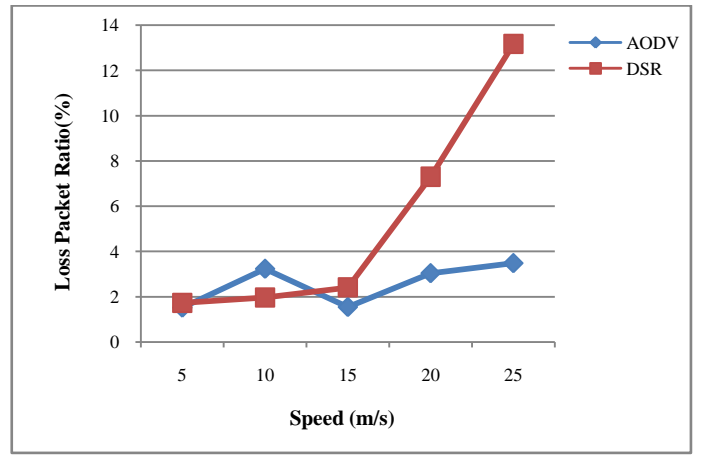

**Fig 24: LPR of 120 nodes using CBR**

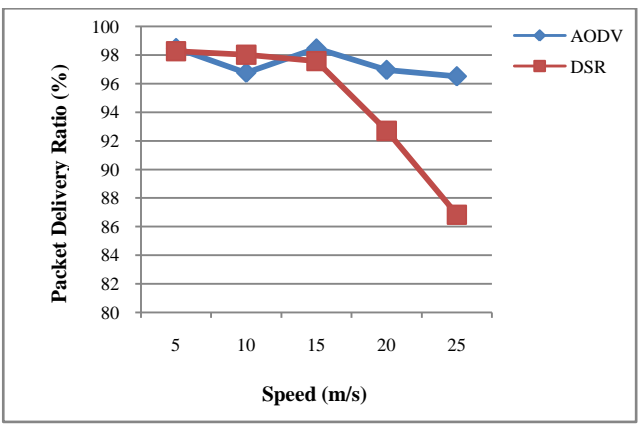

**Fig 22: PDR of 120 nodes using CBR**

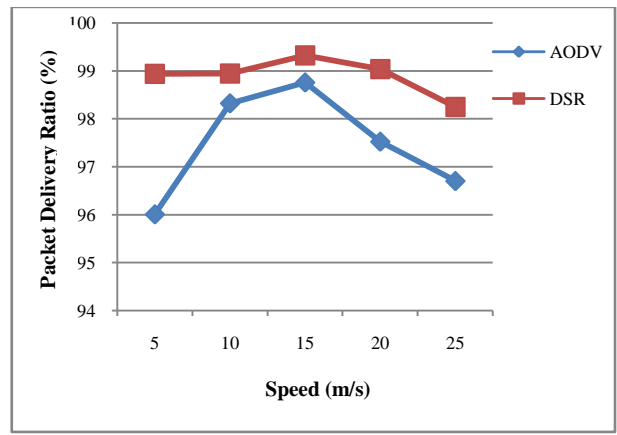

**Fig 25: PDR of 150 nodes using TCP**

35

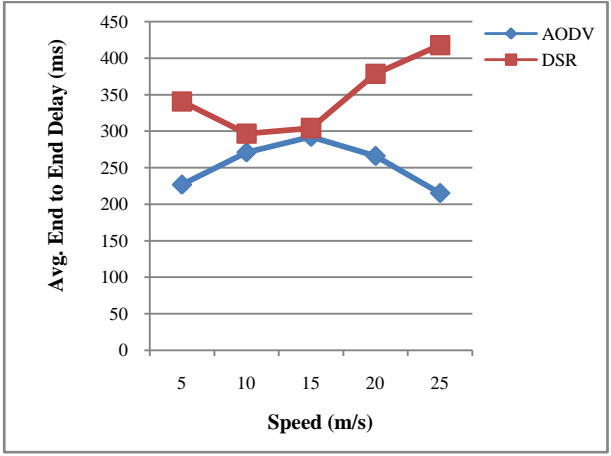

Fig 26: Avg.E-2-E delay of 150 nodes using TCP

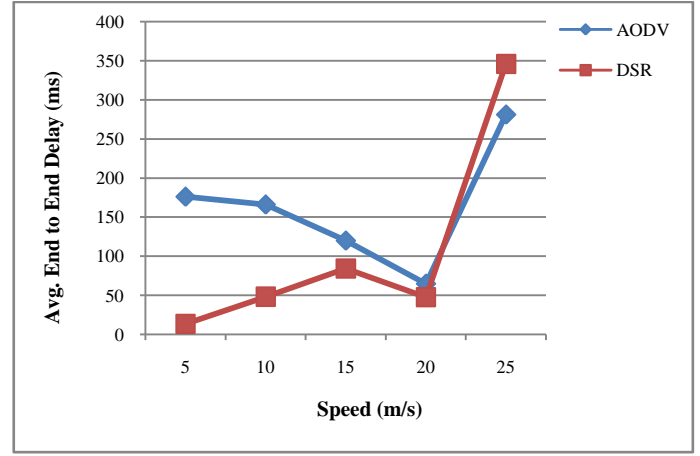

Fig 29: Avg.E-2-E delay of 150 nodes using CBR

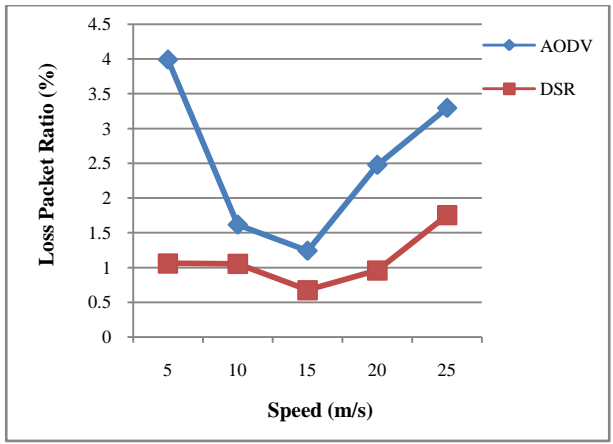

Fig 27: LPR of 150 nodes using TCP

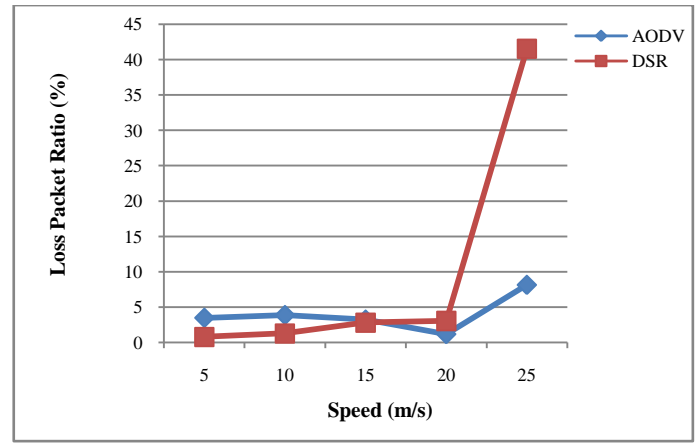

Fig 30: LPR of 150 nodes using CBR

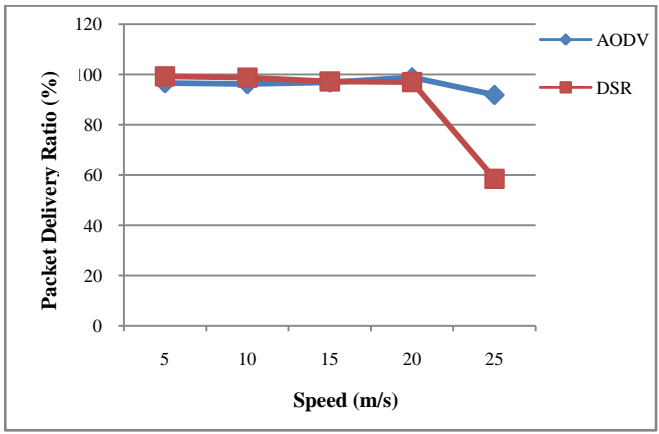

Fig 28: PDR of 150 nodes using CBR

### 5.3 Analysis Table

After analysis of AODV and DSR the results have been shown in a table. We define a standard for simulation results. We consider 30 nodes as low density, 90 nodes as average density and 150 nodes as high density. We also consider 5 m/s as low speed, 15 m/s as average speed and 25 m/s as high speed.

The standard for PDR values (approx.) defines below:
High: >=98%
Average: 96% to 97%
Low: <=95%
The standard for E-to-E values (approx.) defines below:
High: >=150ms
Average: 51 to 150ms
Low: <=50ms
The standard for LPR values (approx.) define below:
High: > 3%
Average: 1.5% to 3%
Low: < 1.5%
Based on our standard we can summarize the following differences between AODV and DSR based on our estimated parameters.

| Nodes Density | Packet Delivery Ratio | | | | Avg.End to End Delay | | | | Loss Packet Ratio | | | |
|---|---|---|---|---|---|---|---|---|---|---|---|---|
| | TCP | | CBR | | TCP | | CBR | | TCP | | CBR | |
| **Low Density** | AODV | DSR | AODV | DSR | AODV | DSR | AODV | DSR | AODV | DSR | AODV | DSR |
| Low Speed | High | High | High | High | High | High | Low | Low | Low | Low | Low | Low |
| Avg.Speed | High | High | High | High | High | High | Avg | High | Low | Low | Low | Avg |
| High Speed | Avg | High | High | Avg | High | High | Avg | Avg | Avg | Low | Avg | Avg |
| **Avg. Density** | | | | | | | | | | | | |
| Low Speed | High | High | High | High | High | High | Avg | Low | Avg | Low | Low | Low |
| Avg.Speed | Avg | Avg | Low | Avg | High | High | Avg | High | High | Avg | High | Avg |
| High Speed | Avg | Avg | Avg | Low | High | High | Avg | High | High | Avg | High | High |
| **HighDensity** | | | | | | | | | | | | |
| Low Speed | Avg | High | Avg | High | High | High | High | Low | High | Low | High | Low |
| Avg.Speed | High | High | Avg | Avg | High | High | Avg | Low | Low | Low | High | Avg |
| High Speed | Avg | High | Low | Low | High | High | High | High | High | Avg | High | High |

## 6. CONCLUSION

This paper illustrates the differences between AODV and DSR based on TCP and CBR connection with various network parameters. In our analytical table we have given our decision based on the graph. This will definitely help to understand the performance of these two routing protocol.

The performance of these two routing protocol shows some differences in low and high node density.

From our experimental analysis we can conclude that in low density with low speed the packet delivery ratio (PDR) of TCP and CBR connection for both protocols is high. In that scenario average end to end delay (E-To-E) is high for TCP connection but low for CBR. The loss packet ratio is low for both routing protocol. If the speed is high the PDR for AODV using TCP is average but high for DSR. For CBR connection result is just opposite for both protocols .E-To-E for TCP is high and low for CBR connection. LPR of AODV using TCP and CBR connection is average. But for DSR using TCP it is low and average for CBR connection.

In high density with low speed, PDR of TCP and CBR connection for AODV is average but high for DSR. Though E-To-E for AODV using TCP and CBR connection is high but it is high in TCP and low in CBR for DSR. LPR is low for DSR and high for AODV using TCP and CBR connection. If the speed is high the PDR for AODV and DSR using CBR is low but using TCP AODV performs average and DSR performs high .E-To-E using TCP and CBR is high for both routing protocol. LPR of AODV using TCP and CBR connection is high .But for DSR using TCP it is average and high for CBR connection.

## 7. REFERENCES

[1]   Ericson, "Communication and Mobility by Cellular Advanced Radio", ComCar project, www.comcar.de, 2002.

[2]   the Internet for ist drive Online, available at: http://www.ist-drive.org/index2.html.

[3]   W. Franz, H. Hartenstein, and M. Mauve, Eds., *Inter-Vehicle-Communications Based on Ad Hoc Networking Principles-The Fleet Net Project.* Karlshue, Germany: Universitatverlag Karlsuhe,November 2005.

[4]   A. Festag, et. al., "NoW-Network on Wheels: Project Objectives,Technology and Achievements", Proceedings of 6th InternationalWorkshop on Intelligent Transportations (WIT), Hamburg, Germany,March 2008.

[5]   Reichardt D., Miglietta M., Moretti L., Morsink P., and Schulz W.,"CarTALK 2000 – safe and comfortable driving based upon inter-vehicle-communication," in Proc. IEEE IV'02.

[6]   Morris R., Jannotti J., Kaashoek F., Li J., Decouto D., "CarNet: A scalable ad hoc wireless network system," 9th ACM SIGOPS European Workshop, Kolding, Denmark, Sept. 2000.

[7]   http://en.wikipedia.org/wiki/List_of_ad_hoc_routing_protocols

[8]   Bijan Paul; Md. Ibrahim; Md. Abu Naser Bikas (*April 2011,* Volume *20– No.3* by IJCA.)"VANET Routing Protocols: Pros and Cons".

[9]   Perkins, C.; Belding-Royer, E.; Das, S. (July 2003)"Ad hoc On-Demand Distance Vector (AODV) Routing".

[10]  Johnson, D. B. and Maltz, D. A. (1996), "Dynamic Source Routing in Ad Hoc Wireless Networks," Mobile Computing, T. Imielinski and H. Korth, Eds., Ch. 5, Kluwer, 1996, pp. 153–81.

[11]  The Network Simulator - ns-2 available at: http://www.isi.edu/nsnam/ns/

[12]  Tutorial for network simulator available at: http://www.isi.edu/nsnam/ns/tutorial/